\documentclass[pre,twocolumn,superscriptaddress,showpacs]{revtex4-1}
\usepackage{graphicx}
\usepackage{color}
\usepackage[usenames,dvipsnames]{xcolor}
\usepackage{amsmath, amsthm, amssymb}
\usepackage{epsfig}
\usepackage{verbatim}
\usepackage{listings}
\usepackage{xspace}
\usepackage{rotating}
\usepackage{natbib}

\newcommand{\be}{\begin{equation}}
\newcommand{\ee}{\end{equation}}	
\newcommand{\bea}{\begin{eqnarray}}
\newcommand{\eea}{\end{eqnarray}}

\begin{document}

\title{Directed Abelian sandpile with multiple downward neighbors}

\author{D. Dhar}
\affiliation{Dept. of Theoretical Physics, Tata Institute of Fundamental Research, Mumbai 400005, India}
\author{G. Pruessner}
\affiliation{Dept. of Mathematics, Imperial College London, London, South Kensington Campus, London SW7 2AZ, UK}
\affiliation{Centre for Complexity Science, Imperial College London, South Kensington Campus, London SW7 2AZ, UK}
\author{P. Expert}
\affiliation{Centre for Complexity Science, Imperial College London, South Kensington Campus, London SW7 2AZ, UK}
\affiliation{Dept. of Physics, Blackett Laboratory, Imperial College London, South Kensington Campus, London SW7 2AZ, UK}
\affiliation{Centre for Neuroimaging Sciences, Institute of Psychiatry, King's College London, London SE5 8AF, UK}
\author{K. Christensen}
\affiliation{Centre for Complexity Science, Imperial College London, South Kensington Campus, London SW7 2AZ, UK}
\affiliation{Dept. of Physics, Blackett Laboratory, Imperial College London, South Kensington Campus, London SW7 2AZ, UK}
\author{N. Zachariou}
\affiliation{Centre for Complexity Science, Imperial College London, South Kensington Campus, London SW7 2AZ, UK}
\affiliation{Dept. of Physics, Blackett Laboratory, Imperial College London, South Kensington Campus, London SW7 2AZ, UK}


\noindent \email{ddhar@theory.tifr.res.in}\\
\email{g.pruessner@imperial.ac.uk}\\
\email{paul.expert@kcl.ac.uk}\\
\email{k.christensen@imperial.ac.uk}\\
\email{nicky.zachariou@imperial.ac.uk}\\

\date{\today}

\begin{abstract}We study the directed Abelian sandpile model on a square lattice, with $K$ downward neighbors per site, $K > 2$. The $K=3$ case  is solved exactly, which extends the earlier known solution for the $K=2$ case. For $K>2$, the avalanche clusters can have holes and side-branches and are thus qualitatively different from the $K=2$ case where avalanche clusters are compact. However, we find that the critical exponents for $K>2$ are identical with those for the $K=2$ case, and the large scale structure of the avalanches for $K>2$  tends to the $K=2$ case.
\end{abstract}


\maketitle

\section{Introduction}
The directed Abelian sandpile model is  a simple variation of the sandpile model first introduced by Bak, Tang and Wiesenfeld \cite{btw}.  It was the first  model of self-organized criticality solved exactly\cite{ddrr89}, and its critical exponents can be determined  in all dimensions $d$. The exponents take the classical values for $d>3$, and for $d=3$, there are logarithmic corrections to power-law behavior\cite{lubeck}. The model is related to other models of non-equilibrium statistical physics like the voter model, Scheideggar river network model, and Takayasu model of aggregating and diffusing particles with injection \cite{dd2006}. It has also found applications  in  more complex situations like modelling economic networks \cite{bak},  growth of droplets of water in  falling rain \cite{andrade} and fracture of ice-sheets \cite{dd2015}.

While the exact solution of the directed Abelian sandpile model on a square lattice with $K=2$ downward neighbors per site is rather elementary, it depends crucially on the fact that avalanche clusters in this case have no holes and thus the problem reduces to that of two annihilating random walkers. If we consider sandpile models with $K > 2$, this property is no longer true, and it is not clear if the problem for $ K > 2$ belong to the same universality class as $K=2$. In fact, direct estimates of critical exponents from Monte Carlo simulations of the model with $K > 2$ show a persistent deviation from the exactly known $K=2$ values \cite{zachariou2015}.

The aim of this paper is to resolve  this discrepancy. We provide an exact solution for the $K=3$ case, and  show that both $ K =2 $ and $K=3$ belong to the same the universality class. While this conclusion is not  very surprising, the exact solution for $K=3$ is of some interest, as we get the exact expression for the  generating function of  the  mean-squared flux at a given depth from the top. This was not done in the earlier study of the $K=2$ case,  where only the critical exponents were deduced using scaling arguments. The $K=3$ avalanches differ  from $K=2$ avalanches mostly near the surface, as can be seen from the pictures of typical avalanche clusters (Fig. 1). We find that the presence of holes and side-branches in the avalanches for $K=3$ is irrelevant in the sense of renormalisation group theory. Finally, we present the results of large scale Monte Carlo studies of this model for $K = 2,3,$ and $4$.   Our data is consistent with all of these being in the same universality class, and the observed deviations of exponents   from the exact theoretical values in earlier simulations may be ascribed to  the effect of significant corrections to scaling. 
\begin{figure*}
	\centering
	\includegraphics[scale=1.6]{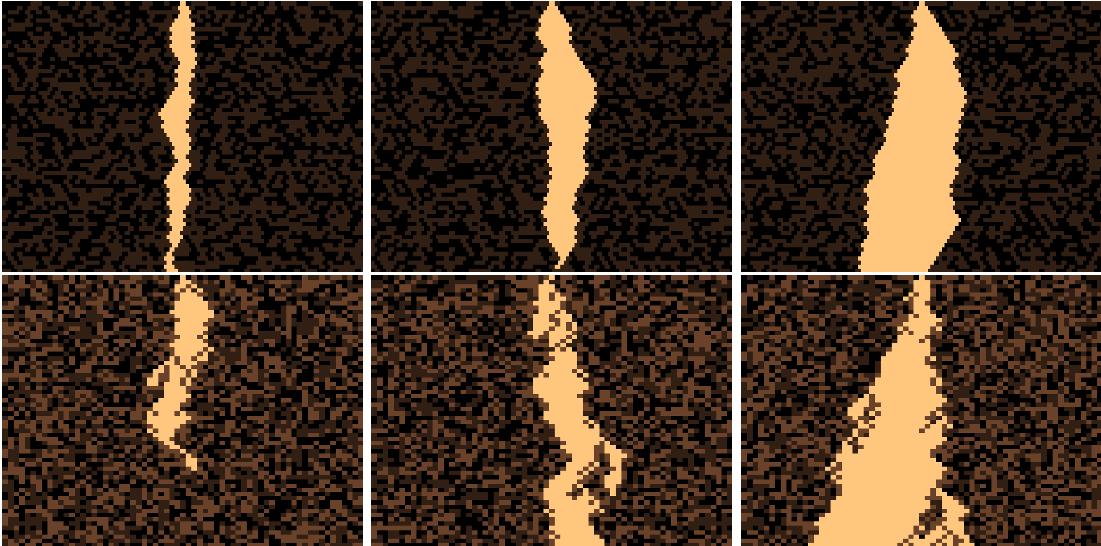}
	\caption{Randomly selected, but representative, examples of avalanches for $K=2$, $s=201, 406,$ and $802$ (top row) and $K=3$, $s=201, 404,$ and $805$ (bottom row).  The height $L$ is $64$ and the width $M$ is $160$; the avalanches have been centred around their seed site and the width shown is $64$ columns for aesthetic considerations. The colour code is lighter color for greater height,  and the sites that have toppled during the avalanche are highlighted in bright copper.}\label{Fig:Avalanches}
\end{figure*}

\section{Definition of the  model}
\begin{figure}
	\includegraphics[scale=0.7]{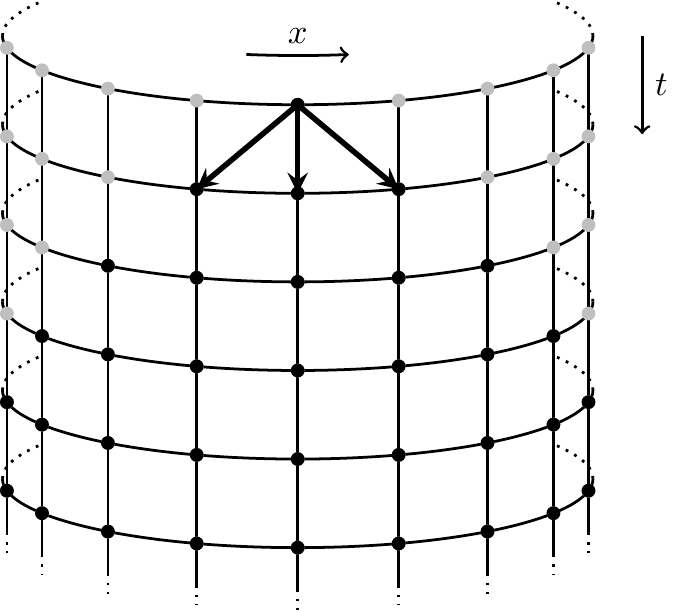}
	\caption{The directed sandpile model on a square lattice with cylindrical boundary conditions and $K=3$ downward neighbours at each site.   If the height at any site exceeds $2$, one particle is transferred to each of the $K$ neighbours in the layer below. The arrows show the directions of particle transfers after toppling at the top site. Filled circles indicate sites in the future light cone of the top site.}
\label{fig:lattice}
\end{figure}

We consider a directed sandpile model on a square lattice.  The lattice is of size (height) $L$ and width $M$. The sites are labelled $\vec{X} \equiv (x,t)$, where $x\in \{0,\ldots,M-1\}$ and the vertical coordinate, thought of as time coordinate, is denoted by $t\in \{0,1,\ldots, L-1\}$. The top row is  $t=0$ and  the $t$-coordinate increases downwards (Fig. \ref{fig:lattice}). We assume periodic boundary conditions in the $x$-direction, so that the $x$-coordinate is defined modulo $M$. 

At each site $\vec{X} \equiv (x,t)$, there is a non-negative integer variable  $h(x,t)$, called the height of the pile.  If $h(x,t) < K$ the site is said to be stable.
If $h(x,t) \geq K$, the site is unstable and is said to topple. When $K=3$, on toppling at $(x,t)$, it will send one particle each to its three downward neighbours $(x-1,t+1), (x,t+1)$ and $(x+1,t+1)$.

Since the avalanche propagation depends only on the layers below the site where the particle is added, without loss of generality, we may assume that  particles are added  only   in the top layer. Thus, we add particles at randomly selected sites in the row $t=0$ and let the configuration relax by toppling. If  a toppling occurs at any site in  the bottom layer  $t=L-1$, three particles are lost from the system. The total number of topplings after adding a particle to the top row is called the avalanche size $s$ and the set of sites where topplings occur, the corresponding  avalanche cluster.

One can easily demonstrate that this is an Abelian\footnote{A model is said to be Abelian if the final state after adding two particles, first at site $(x_1,0)$ and the second at site $(x_2,0)$ does not differ from the final state of the system when the  particles are added in the reverse order.} model. There are $K^{LM}$ stable  recurrent  configurations and all occur with equal probability in the steady state, if particles are added everywhere \cite{dd2006}. However, if particles are added only in the top layer, then the system breaks into $K^{L-1}$ disjoint sectors. In the thermodynamic limit of large lattice, i.e., $L, M \gg 1$, the avalanche statistics in the steady state is independent of the sector.

\section{Calculating the 2-point correlation function}

Consider the pile in the steady state and add a particle at the origin $ \vec{O} \equiv (0,0)$. Let $ \eta(\vec{X})$ denote the indicator variable that this causes a toppling at  the site $\vec{X} \equiv (x,t)$. 
We define the two points correlation function $G_2(x,t)$ as the expected number of topplings at  $\vec{X}$, given that $\vec{O}$  topples after adding a particle at $\vec{O}$, i.e.,
\be \label{eq:1}
G_2(x,t) =  \langle \eta(\vec{X}) \rangle.
\ee
Clearly, $G_2(0,0) =1$ as we are considering the expected no. of topplings under the condition that site $\vec{O}$ topples.
In the steady state, any given site $\vec{X}=(x,t)$ has equal probability of being with height $0,1$ or $2$. Thus, if we focus on the activity in the layer above,  the probability of topplings at $(x,t)$ in a single avalanche equals to 1/3 times the expected number of  upward neighbors  that have toppled in the avalanche. 

Thus, $G_2(x,t)$ satisfies the equation
\begin{equation} \label{eq:2}
G_2(x,t) = \frac{1}{3}\sum_{\delta=-1}^{1} G_2(x + \delta,t-1).
\end{equation}
With the boundary condition $G_2(x,t=0) = \delta_{x,0}$, this equation determines the function $G_2(x,t)$ for all $x$ and $t$.
 
If we define the characteristic function
\be \label{eq:3}
\tilde{G}_2(k,t) = \sum_x G_2(x,t) e^{i k x},
\ee
it is easily seen that 
\be
\tilde{G}_2(k,t) = \left[ \frac{1 + 2 \cos k}{3} \right]^t.
\ee
Substituting this into an inverse transform of Eq.~(\ref{eq:3}), we get
\be
G_2(x,t) = \int_0^{2 \pi} \frac{dk}{2 \pi} \left[ \frac{1 + 2 \cos k}{3} \right]^t e^{-i k x}.
\label{g2eq1}
\ee
Note that $G_2(x,t)$ vanishes strictly for all $|x| > t$. We call the region $|x| \leq t$ the future light cone of the origin, see Fig.~\ref{Fig:Avalanches}. For large $t$, $G_2(x,t)$ is well-approximated by a Gaussian of zero mean and variance $2 t/3$.

We define $\Phi(t)$ as the random variable that measures the  number of topplings in the layer $t$ caused by driving at $\vec{O}$:
\be
\Phi(t) = \sum_x \eta(x,t).
\ee

Then, it is easily seen that 
\be
\langle \Phi(t) \rangle = \sum_x G_2(x,t) = 1, {\rm ~ for ~all~} t.
\label{g2eq1a}
\ee

We would now like to calculate the mean square flux $\langle \Phi^2(t) \rangle$. By definition: 
\be
\langle \Phi^2(t) \rangle = \sum_{x_1,x_2}  G_3(x_1,x_2|t).
\label{g3eq0}
\ee
where 
\be \label{eq:11}
 G_3(x_1,x_2|t) = \langle \eta(x_1,t) \eta(x_2,t) \rangle.
\ee

Note that since each site topples at most once, $G_3(x_1,x_2|t)$ is in fact the probability that sites $\vec{X_1}$ and $\vec{X_2}$ both topple in a given avalanche.

\section{Calculation of the $3$-point function $G_3(x_1,x_2|t)$}

Let us denote $(x_1+\delta_1,t-1)$ and $(x_2+\delta_2,t-1)$, $\delta_1,\delta_2\in\{-1,0,1\}$ to be the upward neighbours of $(x_1,t)$ and $(x_2,t)$, respectively. 

For $x_1 \neq x_2$, if we consider the layer above, since $(x_1,t)$ and $(x_2,t)$ have 3 upward neighbours each, the probability that $(x_1,t)$ or $(x_2,t)$ topples is equal to $1/3$ times the number of topplings of their upward neighbours. Hence,
\be
 G_3(x_1,x_2|t) \!=\! \frac{1}{9}\!\sum_{\delta_1,\delta_2 =-1}^{+1} \!\!\langle \eta(x_1\!+\!\delta_1,t\!-\!1) \eta(x_2\!+\!\delta_2,t\!-\!1) \rangle.
\label{eqg3}
\ee
Thus, from Eq.~\eqref{eq:11} we get  for  all $x_1 \neq x_2$
\be
G_3(x_1, x_2|t) = \frac{1}{9}\sum_{\delta_1,\delta_2 =-1}^{+1} G_3(x_1 +\delta_1,x_2 + \delta_2|t-1).
\label{g3eq1}
\ee
Note that for the special casse of $x_1 = x_2$, since $\eta^2(x_1,t)= \eta(x_1,t)$,
we have
\be
G_3(x_1,x_1|t) = G_2(x_1,t) \quad \forall x_1.
\label{g3eq2}
\ee

It is easy to verify that 
\be \label{eq:17}
G_3(x_1,x_2|t) = G_2[\vec{X_1}|\vec{O}] ~ G_2[\vec{X_2}|\vec{O}]
\ee
satisfies the linear equation (\ref{eqg3}), where   $G_2[\vec{X}|\vec{Y}]$ denotes  the expected number of topplings at $\vec{X}=(x,t)$, in an avalanche generated by a  particle addition at $\vec{Y}=(y,t')$.  Clearly,
\be
G_2[\vec{X}|\vec{Y}] = G_2(\vec{X} -\vec{Y}) = G_2(x-y,t-t').
\ee
The function $G_2[\vec{X}|\vec{Y}]$ inherits the light-cone structure of $G_2(x,t)$:  it is nonzero only if $\vec{X}$ is in the future-light cone of $\vec{Y}$.

To find a solution that is also consistent with the boundary conditions Eq. (\ref{g3eq2}), we consider the following superposition of such solutions \cite{ddrr89}:
\be 
G_3(x_1,x_2|t) = \sum_{\vec{Y}} f(\vec{Y}) G_2[\vec{X_1}|\vec{Y}]  G_2[\vec{X_2}|\vec{Y}].
\label{g3eq3}
\ee
Here the summation over $\vec{Y}$ could extend over all sites. However, due to the light-cone structure of the propagator $G_2$, only the sites which are in the future light-cone of $\vec{O}$ and in the past light-cones of both $\vec{X_1}$ and $\vec{X_2}$ contribute to the summation.  

Then, for $\vec{X_1}=\vec{X_2}$ the equations determining the unknown coefficients $f(\vec{Y})$ are the boundary conditions Eq.~(\ref{g3eq2}), which become
\be
G_2(\vec{X}) = \sum_{\vec{Y}} f(\vec{Y}) G_2^2 [\vec{X}|\vec{Y}], \quad \forall \vec{X},
\label{eqf}
\ee
where the summation over $\vec{Y}$ is over all sites that are in the future light-cone of $\vec{O}$, and the backward light-cone of $\vec{X}$.

These equations are coupled linear equations which can be used to determine the unknown function $f(\vec{Y})$ at 
sites $\vec{Y}$ in the past-light cone of $\vec{X}$, and hence can be determined recursively starting from the driving site.
 
In fact, we do not need to know  the $f(\vec{Y})$'s for all $\vec{Y}$; the knowledge of their sum for each constant-time layer is sufficient. Let us denote this sum, for all times $t \geq 0$, as
\be \label{eq:24}
F(t) = \sum_x f((x,t)).
\ee

Now, if we substitute our general solution from Eq.~(\ref{g3eq3}) into our expression for $\langle \Phi^2(t)\rangle$ as stated in Eq.~(\ref{g3eq0}) we get:
\be
\langle \Phi^2(t) \rangle = \sum_{x_1,x_2}\sum_{\vec{Y}} f(\vec{Y}) G_2[\vec{X_1}|\vec{Y}]  G_2[\vec{X_2}|\vec{Y}].
\label{eq:25}
\ee

Doing summations over $x_1$, $x_2$ and $t'$,  using Eq. (\ref{g2eq1a}),  and (\ref{eq:24}), we get 
\be
\langle \Phi^2(t)\rangle = \sum_{t'=0}^{t}  F(t').
\ee

Now let us define
\be \label{eq:28}
K(t) = \sum_{x} G_2^2(x,t)
\ee
and note that $K(t)$ vanishes with $G_2(x,t)$ when $t<0$.

Then, summing over different sites $\vec{X}$ in the layer $t$, in Eq.~(\ref{eqf}), and using Eq.~(\ref{g2eq1a}) yields
\be \label{eq:29}
1 = \sum_{t'=0}^{t} F(t') K(t-t').
\ee

This equation differs from that in \cite{ddrr89} by a factor, due to different choice of normalization of $G_2(\vec{X})$ here. Also, note that  the summation over $t'$ may be extended  to $+\infty$, as  $K(t-t')$ vanishes with $t<t'$.

Now we define
\begin{subequations}
\begin{align}
\tilde{F}(z) &= \sum_{t=0}^{\infty} F(t) z^t \\
\tilde{K}(z) &= \sum_{t=0}^{\infty} K(t) z^t
\end{align}
\end{subequations}
to be the generating functions of, ${K}(z)$ and ${F}(z)$, respectively. In terms of these generating functions, Eq.~\eqref{eq:29} can be expressed as
\be
\frac{1}{1-z} = \tilde{K}(z) \tilde{F}(z).
\label{g3eq4}
\ee

Now let us remind ourselves that $G_2(x,t)$ is the expected number of topplings at site $(x,t)$ given that there was a toppling at the origin $\vec{O}$. A site can topple at most once, hence its expected number of topplings is exactly its probability to topple. Also, based on the Abelian property we know that the total number of topplings at another site $(x',t')$ is the sum of the topplings triggered by $(x,t)$ toppling. Hence, we can write the expected number of topplings at site $\vec{X'}=(x',t')$ as
\begin{align}
G_2[\vec{X'}] &\equiv G_2(x',t')  = \sum_{x}   G_2(x,t) G_2(x'-x,t'-t) \notag \\
&\equiv \sum_{x}   G_2[\vec{X}]G_0[\vec{X'}|\vec{X}]
\end{align}
for $0\leq t\leq t'$. But then, the expected number of topplings at site $(0,2t)$ is just 
\begin{equation}
G_2(0,2t)  = \sum_{x}   G_2[(0,2t)|(x,t)] G_2[(x,t)|(0,0)]
\end{equation}
Also,  $G_2[(0,2t)|(x,t)] = G_2[(-x,t)|(0,0)] = G_2(x,t)$. 
Substituting this in Eq.~\eqref{eq:28} yields
\be
K(t) = G_2(0,2t).
\ee 
Thus, the generating function becomes
\be
\tilde{K}(z)= \sum_{m=0}^{\infty}  G_2(x=0,t=2m) z^m.
\ee

Now let us define $H(z) = \sum\limits_{m=0}^{\infty} G_2(0,m) z^{m}$ and hence
\be
H(z)+H(-z)=2\sum_{m=0}^{\infty} G_2(0,2m) z^{2m}.
\ee
Since $\tilde{K}(z)$ is sum only over even values of $t$ of $G_2(0,t)$, we have
\be \label{eq:36}
\tilde{K}(z) = \frac{1}{2}[ H(\sqrt{z}) +  H(- \sqrt{z}) ].
\ee
Substituting Eq.~(\ref{g2eq1}) with $x=0$ into our definition of $H(z)$ and evaluating the geometric sum yields
\be
H(z) = \int_0^{2 \pi} \frac{dk}{2 \pi}  ~\frac{3}{[3-z( 1 + 2 \cos k)]}.
\ee
Finally, evaluating this elementary integrals yields
\be \label{g2eq2}
H(z) = \frac{3}{\sqrt{(1 -z)(3 + z)}}.
\ee
We can substitute this into, using Eq.(\ref{eq:36}), to get
\be
\tilde{K}(z) = \frac{3/2}{\sqrt{(1\!-\! \sqrt{z})(3\!+\! \sqrt{z})}} + \frac{3/2}{\sqrt{(1 \!+\! \sqrt{z})(3\!-\! \sqrt{z})}}.
\label{eqK}
\ee
One can substituting this in Eq.\eqref{g3eq4}, to get $\tilde{F}(z)$ as an explicit function of $z$.  Note that odd powers of $\sqrt{z}$ will cancel out in the expansion.

From the fact that the dominant singularity of $\tilde{K}(z)$ for $z$ near $1$, is of the form $( 1-z)^{-1/2}$, we see that for $z$ tending to $1$ from below, the
leading behavior of $\tilde{F}(z)$ also is $(1 -z)^{-1/2}$.  Hence $F(t)$ varies as $t^{-1/2}$ for large $t$ and then $\langle \Phi^2(t) \rangle$ varies as $t^{1/2}$.  
This is the same behavior as found for the case $K=2$ in \cite{ddrr89}.  The probability that $\Phi(t)$ is not zero decreases as $t^{-1/2}$ for large $t$. But once it is non-zero, its typical value is of order $t^{+1/2}$, consistent with the mean value $1$, see Eq.\eqref{g2eq1a}.  Then the mean value $\langle \Phi^2(t) \rangle$ would be expected to grow as $t^{1/2}$ for large $t$.

\begin{figure}
	\includegraphics[scale=0.4]{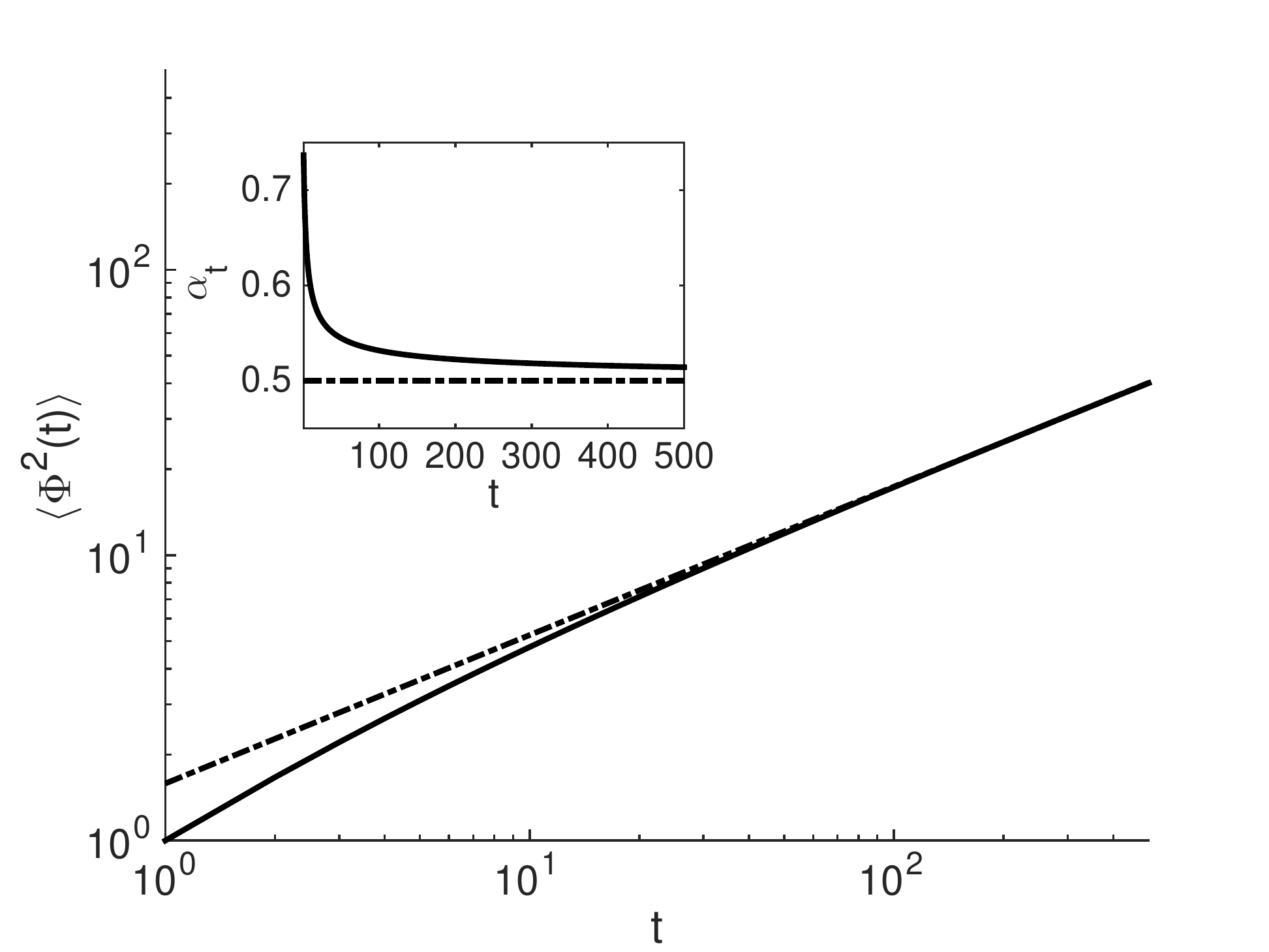}
	\caption{The main figure shows the behavior of the mean squared flux $\langle \Phi^2(t) \rangle$ as a function of the depth $t$, $1\leq t\leq 500$, determined from the exact series expansion of $\tilde{F}(z)$ (full line). The best visual fit to the function $f(t)=a t^{\alpha}$ given by the estimated parameters $a \approx 1.58, \alpha \approx 0.52$ (dashed-dotted line). In the inset, we plot effective exponent $\alpha_{t}$, defined in eq.~(\ref{eq:alphat}) (solid line) which slowly converges to the exact value 0.5 (dashed-dotted line), as $\langle \Phi^2(t) \rangle\propto t^{1/2}$. Both plots in this figure show that corrections to scaling are still large for avalanches of duration of order $5\cdot 10^2$.}
\label{Fig:3}
\end{figure}

We now illustrate in Fig.~\ref{Fig:3} the importance of taking into account the corrections to scaling in estimating exponents from numerical data. First, we plot the exact values of $\langle \Phi^{2}(t)\rangle$ for $1\leq t\leq 500$. These values were determined using Eqs.~(\ref{g3eq4}) and (\ref{eqK}) to expand $\tilde{F}(z)$ as a Taylor series in $z$ with Mathematica. A simple visual fit to a power-law gives $\langle \Phi^{2}(t) \rangle\approx at^{\alpha}$, with $a\approx 1.58$ and $\alpha\approx 0.52$. Secondly, we plot  the effective exponent $\alpha_{t}$. $\alpha_{t}$ is defined in terms of the exact values of $\langle \Phi^{2}(t) \rangle$ at $t$ and $t+1$ by:
\begin{equation}
	\alpha_{t}=\frac{\log\left(\langle \Phi^{2}(t+1) \rangle/\langle \Phi^{2}(t) \rangle\right)}{\log\left((t+1)/t\right)}.\label{eq:alphat}
\end{equation}
We see that 	the effective exponent converges very slowly to the exact value 0.5. These two plots show that corrections to scaling are rather large in this problem.

Conversely, knowing that $\langle \Phi^{2}(t)\rangle$ varies as $t^{1/2}$ implies that the probability that an avalanche has duration greater than $t$ goes as $t^{-1/2}$. It follows that the avalanche duration exponent is $3/2$. The avalanche dimension $D$ in a directed model is identical to the duration exponent, that is $D=3/2$ (Sec. 8.4.3 in \cite{Pruessner:2012:Book}). All other exponents follow and we can verify that they are the same as the case of coordination number $K=2$.

We note that a similar analysis has been reported for a directed sandpile model in \cite{andrade}. In this paper, the authors determine the exact two-point correlation function $G_2(x,t)$ for a model with the following rules: the critical height is $4$, and on toppling at $(x,t)$, a particle is transferred to $(x-1,t+1)$ and $(x+1,t+1)$, and two particles to $(x,t+1)$. With these rules, the functions $K(t)$ and $F(t)$ were determined recursively numerically for  small $t$ ($t < 500$), and it was found that the corrections to scaling are large. However, no explicit analytical expressions for $F(t)$ or $K(t)$ were found.

\newcommand{\ie}{\emph{i.e.}\xspace}
\newcommand{\elabel}[1]{\label{eq:#1}}
\newcommand{\eref}[1]{(\ref{eq:#1})}
\newcommand{\Eref}[1]{Eq.~(\ref{eq:#1})}

\newcommand{\slabel}[1]{\label{sec:#1}}
\newcommand{\sref}[1]{Sec.~\ref{sec:#1}}
\newcommand{\Sref}[1]{Section~\ref{sec:#1}}

\newcommand{\tlabel}[1]{\label{tab:#1}}
\newcommand{\tref}[1]{tab.~\ref{tab:#1}}
\newcommand{\Tref}[1]{Tab.~\ref{tab:#1}}

\newcommand{\flabel}[1]{\label{fig:#1}}
\newcommand{\fref}[1]{Fig.~\ref{fig:#1}}
\newcommand{\Fref}[1]{Figure~\ref{fig:#1}}
\newcommand{\subfref}[1]{\subref{fig:#1}}

\newcommand{\ave}[1]{\left\langle #1\right\rangle}

\newcommand{\gcomment}[1]{\textcolor{blue}{\bf GP: #1}}


\section{Numerical simulations}
The directed sandpile is comparatively easy to implement and study numerically even on very big lattices, because avalanches progress in one direction only, \ie, no backward avalanches \cite{KadanoffETAL:1989} or multiple topplings occur \cite{PaczuskiBassler:2000}. The aim of the numerics is to provide reliable numerical estimates of (supposedly universal) critical exponents. In the following we will discuss pertinent issues in relation to the numerical simulations, the fitting models used and the critical exponent estimates found numerically.

\subsection{Initialisation}
 
The directed sandpile is deterministic up to the driving in the first
row. As every state is recurrent, one might naively start
from an empty lattice $h(x,t)=0$ as this indeed belongs to the stationary state. Because every toppling moves $K$ particles downstream and every particle
added performs $L$ moves before leaving the system, $K$ particles additions cause $L$ topplings (and thus $KL$ moves)
on average, \ie, the average avalanche size is 
\begin{equation}
\ave{s}=L/K,
\label{ave_s}
\end{equation}
where avalanche sizes $s$ are measured as the number of topplings that occur in the system after a
driving attempt (deposition of a particle in the top row). Note that this definition
includes avalanche size $s=0$.

Starting from an empty initial configuration, the first moment of the cluster size $s$ in a
system with $K=4$ and $L=512$ (with $M=3072$, see below) shows very good convergence within less than
$2\cdot10^7$ avalanching attempts (\ie, particles deposited in the top
row). 
However, the second moment of $s$ still shows signs of drift after
$5\cdot10^{10}$ avalanches.
 Higher moments show similar long  transients, but are more noisy. This only underlines the
fact that the steady state of model studied here has very slowly decaying temporal correlations, even though it has no spatial correlations.

To avoid these lengthy transients, we resorted to initialisation with
random, independently, uniformly distributed $h(x,t)<K$. As the system
is then no longer forced into an exceptional initial state, there is no
noticeable drift in {\em any} of the moments measured. We have verified numerically 
for smaller lattices that random initial states generate the same moment estimates as starting from an empty system and taking estimates after those very long transients. 

\subsection{Parameters and Results}
\begin{figure}
	\includegraphics[scale=0.4,angle=90]{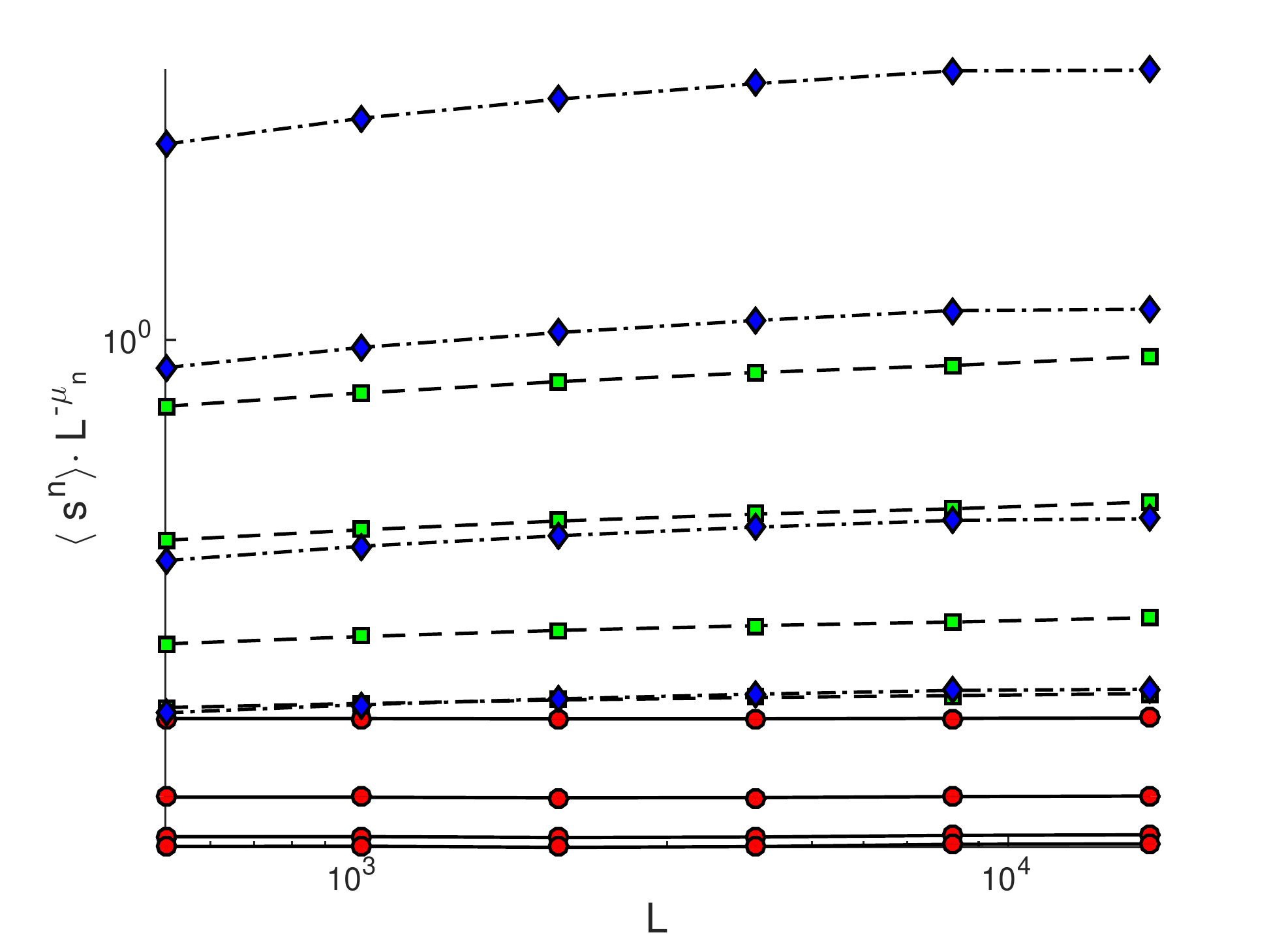}
	\caption{Plot of scaled moments $a_n L^{(3n-1)/2}$ versus the height of the lattice $L$ for $n = 2,3,4,5$. The data shown is for different  $K : ~ 2$ (red circles), 3 (green squares) and 4 (blue diamonds). For the same $K$, a higher curve corresponds to higher $n$. }
\label{Fig:4}
\end{figure}

We used systems with sizes  $L = 2^r$, with $r$ taking integer values from $4$ to $14$,  and    $K=2,3,4$ and
widths $M \geq L(K-1)$ so that even the largest avalanche possible cannot topple all sites of a row.
We used the Mersenne Twister pseudo-random number generator \cite{MatsumotoNishimura:1998a}, which is (after initialising randomly the lattice) needed only
to determine the site in the top row that receives a particle from the
external drive. In order to determine scaling exponents, we measured
moments $\ave{s^n}$ of the avalanche size $s$ for $n=2$ to $5$ (Fig. \ref{Fig:4}).

Statistical errors were determined by accumulating data over chunks of
$10^6$ avalanches and measuring the variance of those estimates. We
produced usually at least several thousand chunks, except for the largest system sizes. If chunks are independent, then the variance does not vary noticeably when chunks are merged. Using that as an indicator, we merged chunks until they
became independent. Based on the now independent measurements,
statistical errors of the mean (across chunks) of a moment are given by
the estimated standard deviation of the moments (across chunks), divided by the square root of the number of independent chunks. 

Because high moments draw most of their weight from very large and thus very rare events, their relative error grows with their order. For example, for $K=3$ and $L = 16384$, averaging over  $3.8 \cdot 10^8$ avalanches, the fractional error in the eighth moment is  approximately $0.014$, while for the second moment it was approximately $.00016$. We report here our results only for moments up to order $5$.  

Standard finite-size scaling of the probability density of the avalanche sizes $s$ implies that the moments scale to leading order in the system size $L$ like
\cite{Pruessner:2012:Book}
\begin{equation}
\ave{s^n} = a_n L^{\mu_n} \quad \text{for $L \gg 1$},
\label{std_scaling}
\end{equation}
where $a_n$ are metric factors,  and $\mu_n = D(1 -\tau+n)$. Here $D$ is the fractal dimension of avalanche clusters, and $\tau\ge1$ \cite{ChristensenETAL:2008}. The critical exponents $D$ and $\tau$ are expected
to be universal, but they are not independent as $\ave{s} = L/K$, Eq.\eqref{ave_s} implies that $D(2-\tau)=1$.

For $\tau>1$ (which is expected in the present case)
the scaling of the variance of the $n$th moment is dominated by that of
the scaling of the $2n$th moment and thus the relative error of
\emph{independent samples} scales like 
\begin{equation}
\frac{\sqrt{\ave{s^{2n}}}}{\ave{s^n}} \propto L^{D(1-\tau)/2}
\end{equation}
and thus grows, independent of the order, with the system size (height).
Meaningful estimates of moments thus require larger and larger sample
sizes for larger systems.

However, simulation time per avalanche grows essentially like the average
avalanche size, which is linear in the system size $L$.
Worse, correlation times grow
with the system size $L$, so that the increased demand on the sample size
for larger system is met with highly increased costs for independent samples.

Using random initialisation, we were able to skip the transient as described above and thus
could produce very large samples even for large system sizes. 
For the smaller system sizes, our sample sizes were  comparatively large,
typically about $10^{10}$ avalanches and more. 
However, to fit well the  accurate estimates for the moments for small $L$, we need to have a large number of corrections to scaling terms, which complicates the analysis. 
Therefore, we excluded system sizes smaller than $L=512$ from further analysis.
\begin{table}
\begin{tabular}{lllll}
$K$ & $D$ & $\tau$  & $D$ (ctos) & $\tau$ (ctos)\\
2 & 1.4999(11) & 1.3333(05) & &  \\
3 & 1.5141(38) & 1.3342(17) & 1.5020(20) & 1.3395(09)\\
4 & 1.5244(78) & 1.3440(34) & 1.4994(35) & 1.3331(23)\\
\end{tabular}
\caption{Numerical estimates of the avalanche dimension $D$, and $\tau$
derived from it via the scaling relation $\tau=2-1/D$. The numerics for $K=2$ were performed
as a reference, as $D=3/2$ and hence $\tau=4/3$ was already known analytically \cite{ddrr89}.
The first two columns are fits without including corrections to scaling. The last two columns
list the exponents extracted when the fit includes corrections to scaling (ctos) Eq. (\ref{CorrectionScaling}) for systems with $K=3$ and $4$. The errorbars listed in the brackets correspond to three standard deviations of the last two significant digits.}\label{expos_from_numerics}
\end{table}

First, the critical exponents $D$ are extracted by applying moments analysis without including
corrections to scaling. In the limit of large system sizes,  the $n$th moment scales according to Eq. (\ref{std_scaling}).
Hence, plotting the measured moments $\ave{s^n}$ against system size $L$
yields estimates for $\mu_n$. According to Eq.\eqref{std_scaling} and the scaling relation
$\tau = 2 - 1/D$, we expect $\mu_n = 1 + D(n-1)$, so we can
extract the critical exponent $D$ by plotting $\mu_n$ vs. $n$. Then the avalanche
size exponent is calculated form the  scaling relation $\tau = 2-1/D$.
The critical exponents are listed in columns 1 and 2 in Tab. \ref{expos_from_numerics}.
The critical exponents we find for $K=2$ are consistent with the theoretical values
$D = 3/2$ and $\tau = 4/3$. However, the critical exponents for $K = 3$ and $4$  show significant deviation  from the  theoretical values and we notice that the deviation
increases with $K$.

However, we now include two corrections to scaling terms in our fitting model 
\begin{equation}
\ave{s^n} = a_n L^{\mu_n} (1+b_n L^{-1/2} + c_n L^{-1})\label{CorrectionScaling}
\end{equation}
and fit each moment vs. system size $L$, using the Levenberg-Marquardt method \cite{PressETAL:1992}. Where convergence was a problem, we provided
initial guesses by fitting first using fewer parameters (which typically
results in sub-standard goodness of fit).
The resulting estimates for exponents $\mu_n$ were accepted for
further analysis if the goodness of fit was at least $0.25$
\cite{PressETAL:1992}. This was the case for $n$ up to $5$. 
One exponent, $\mu_1$, is known to be unity from
Eq.\eqref{ave_s}, which means that 
$\mu_n$ needs to be fitted
linearly against $1+D(n-1)$ for $n=2,3,4,5$. Because the $\mu_n$ are, for different $n$, all based on the
same data (the set of avalanche sizes generated), they are bound to be correlated, but this is difficult to 
quantitfy reliably, except by using a rather brutal upper bound. Taking that course of action, we have effectively assumed
that the estimates of $\mu_n$ for $n=2,3,4,5$ may have been derived from \emph{distinct} samples, by multiplying 
each error bar by a factor $\sqrt{4}$.
The resulting estimates for the exponent
$D$ (and implicitly for $\tau$) are shown in Tab.~\ref{expos_from_numerics}, columns 3 and 4, respectively.

These estimates fit acceptably well with the theoretical value of
$D=3/2$ and $\tau=4/3$, but only for very large sample sizes, system sizes and CPU time.

\subsection{Summary and Conclusions}
We have studied the directed Abelian sandpile model on a square lattice
with $K$ downward neighbors. When $K = 2$, avalanche clusters are compact without any holes. Using this property, the $K =2$ case has previously been solved exactly \cite{ddrr89}. When $K > 2$, avalanches clusters typically contain holes, that is, they are no longer compact. Hence, the previous
derivation for the $K=2$ case cannot be extended to the $K>2$ case and it is
an interesting question whether $K > 2$ belong to the same universality class
as $K=2$.

In this paper, we calculated exactly the exponents for the $K=3$ case, where the problem is complicated by the fact that  avalanche clusters are no longer compact.  We find that the critical exponents are identical to the $K=2$ case, that is, the avalanche dimension $D=3/2$ and the avalanche size exponent $\tau = 4/3$. In addition, we get exact expression for other observables, for example,  generating function $\tilde{F}(z)$ of  the mean-square flux.  Our result shows that the deviations from $K=2$ values observed in  recent numerical studies are due to corrections to scaling.

We performed large scale numerical simulations of the generalized directed Abelian sandpile model for $K = 2, 3$ and $4$. Although the empty state is a recurrent state, it is not a typical state. Initialising the system in the empty state results in extremely long transients before correlations caused by this exceptional state vanish. Initialising the system in a random recurrent state minimised the transient time before avalanche size moments estimates no longer drifted.

Using system sizes $L = 512, 1024, \ldots, 16384$, moments analysis yields numerical estimates for the avalanche dimension $D$ and hence $\tau = 2 - 1/D$. For $K = 2$, the numerical estimates are consistent with the exact result. However, for $K = 3$ and $4$, there are large corrections to scaling effects. If these are taken into account (see Eq.\eqref{CorrectionScaling}), the resulting numerical estimates for the critical exponents are consistent with the exact findings, that is, the directed Abelian sandpile model belong to the same universality class for all $K \geq 2$.

\subsection{Acknowledgements} P.E. acknowledges financial support from a PET methodology programme grant from MRCUK (ref no. G1100809/1).  DD's research  is supported partially by the Indian DST  via grant DST-SR/S2/JCB-24/2005.

\end{document}